\documentclass[%
 reprint,
 amsmath,amssymb,
 aps,
floatfix,
]{revtex4-2}
\usepackage{amsmath}
\usepackage{amssymb}
\usepackage{chemformula} % optional, but useful for chemistry
\usepackage{hyperref}
\usepackage{comment}
\usepackage{booktabs}
\usepackage{siunitx}
\usepackage{amsmath}
\usepackage{graphicx}% Include figure files
\usepackage{dcolumn}% Align table columns on decimal point
\usepackage{bm}% bold math
\usepackage[utf8]{inputenc}

\usepackage[T1]{fontenc}

\begin{document}

\preprint{APS/123-QED}

\title{\Large Supernova Neutrinos and the Origin of Biomolecular Homochirality}

\author{Amirmasoud Jannat}
\author{Soroush Shakeri}
\author{Farhad Shahbazi}
\affiliation{Department of Physics, Isfahan University of Technology, Isfahan 84156-83111, Iran}

\date{\today}% It is always \today, today,

\begin{abstract}
We investigate the role of parity-violating interactions between supernova neutrinos and chiral molecules in nearby interstellar molecular clouds as a potential source of biomolecular homochirality.
We introduce neutrino interactions into the autocatalytic chemical  reactions  in a far-from-equilibrium  noise-induced system. These interactions create a directional bias between L and D enantiomers in  the racemization reactions, which is amplified by autocatalysis and stochastic fluctuations. We solve  the stochastic equations within the Itô sense to obtain the dynamics of the  probability distribution of the  enantiomeric excesses,  offering an astrophysical scenario for the delivery of   homochirality seeds to Earth by meteorites. In spite of the weak interactions of supernova neutrinos, our framework introduces an amplification mechanism to yield a  considerable enantiomeric excess of more than $10\%$,  in agreement with the latest chemical analysis of the meteorites. Moreover, we scan over the parameter space of the model, inferring from the observational values in order to explore the window of opportunity  to generate the initial seeds of homochiral states in interstellar molecular clouds.
\end{abstract}
\maketitle

\section{\label{sec:level1}{Introduction}}
“\textit{\bf What is life?}” 
As Schrödinger said, "Life seems to be orderly and lawful behaviour of matter, not based exclusively on its tendency to go over from order to disorder, but based partly on existing order that is kept up \cite{Schrodinger1944}". 
While no universally accepted definition of life can separate living things from nonliving ones, one step toward better understanding life is to identify the common characteristics of living systems \cite{Carroll2016}. Life, as currently understood, has some key characteristics: order, evolutionary adaptation, regulation, energy processing, growth and development, response to the environment, and reproduction \cite{shen2020campbell}. 

However, there are other common features among living systems, such as biomolecular homochirality, which was discovered by Pasteur in 1848 \cite{Pasteur1848}. In fact, living systems exhibit homochirality, with predominantly left-handed amino acids and right-handed sugars \cite{RN1,RN6,RN16,RN17,salam1991role,RN19,RN87,RN46,RN45,RN47,RN16,RN72}.  There are different approaches to explain the origin of biomolecular homochirality; this could be generated by chemical processes such as auto-catalytic chemical cycles or crystallization in far-from-equilibrium conditions due to both kinetics and thermodynamics of the system \cite{Blackmond2010CSH,Blackmond2004Nature, RN87, HiggsBlackmond2025,ozturk2023origin}. There are also models which propose that fundamental parity-violating weak interactions can trigger this phenomenon \cite{RN72,RN3,RN101, hegstrom1980calculation,laerdahl1999fully, zel1977energy,rein1979parity,quack1986measurement,eills2017measuring, flambaum2025enhancement}. There is also some evidence that coupling of electron spin and molecular chirality could trigger an achiral magnetized surface that acts as a chiral agent via spin-controlled asymmetric interactions, known as Chiral-induced Spin Selectivity \cite{ozturk2025life}.

 It has been shown that, under some specific conditions, autocatalytic cycles could break chiral symmetry, leading to a homochiral final state \cite{Frank1953}. Note that there are various chemical schemes for this process \cite{Calvin1969,Kondepudi2001,Islas2005,Kondepudi1985,Iwamoto2003}, but as far as experimental results are concerned, there is only one chemical scheme,  based on dimerization of enantiomers in an autocatalytic cycle \cite{HiggsBlackmond2025}. In these models, noise can  be introduced in far-from-equilibrium systems, which is taken as the origin of biomolecular homochirality \cite{RN70}.  
A chiral molecule could be modeled as a symmetric double-well potential of two enantiomeric states with the same energies, and a weak parity-violating interaction breaks this symmetry and induces a small energy difference between the two  enantiomeric states.

Parity Violation Energy Difference (PVED) makes one of the enantiomers more stable and preferable to build life blocks \cite{RN1}. Parity violation could happen by Circularly Polarized Light (CPL), or weak interactions \cite{RN71}. There are various particles, whether in standard model (SM) of particle physics or beyond that such as SM neutrinos \cite{RN17,RN18,RN19}, dark matter candidates specially sterile neutrinos  and  Axion Like Particles (ALPs) \cite{RN27,RN25}, and Weakly Interaction Massive Particles (WIMPs) \cite{RN24} which can induce parity violating interactions among  biomolecules.

The molecules of life could be formed on  Earth, or they could have an extraterrestrial origin. In support of the latter idea, complex molecules might be created initially in the interstellar medium or in an appropriate situation \cite{McGuire2022} triggered by an astrophysical event. Then some meteorites which contain enantiomeric excess might carry them  onto Earth as the seeds of life \cite{Boyd2011,RN22,Lawless1971,cronin1971amino,Callahan2011}. In an alternative scenario, the solar system could cross an Interstellar Molecular Cloud (IMC). It is argued that the period of life that appeared on  Earth is estimated to be between 3.8 and 3.5 billion years ago; this provides a small window of 300 million years (Myr) for the emergence of life \cite{cline1996homochiral,cline2005physical}. Also, there are some proposals in which the solar system was formed by an IMC triggered by a supernova \cite{cameron1977supernova, banerjee2016evidence}; therefore, it is plausible that, during the formation of Earth, enantiomeric excess of chiral molecules  pre-existed. Some of the meteorites showed enantiomeric excess in amino acids that are utilized in living systems, such as Murchison with up to 10\% enantiomeric excess \cite{RN22, mcguire2016mirror}. It is worth mentioning to note that the trace of the chiral molecule propylene oxide $CH_3CHOCH_2$ has been discovered in the interstellar medium \cite{2016Sci...352.1449M}.
Additionally, a recent study provides evidence of a protein-templated mechanism for sequence-specific DNA synthesis \cite{deng2026protein}, one could propose a scenario for the origin of life similar to RNA world, but instead of nucleotide bases and RNAs, one can consider proteins crystallized in a bath of amino acids with enantiomeric excess to explain the origin of life, since proteins now could be considered as agents for both metabolism and replication processes.

Some models have attempted to explain the extraterrestrial origin of biomolecular homochirality using parity-violating interactions and the subsequent amplification in autocatalytic processes.  It has been proposed  that the chirality of the amino acids could be generated
in the magnetic field of a newborn neutron star after supernova  via parity-violating interaction of neutrinos with amino acids. In the Supernova Neutrino Amino Acid Processing (SNAAP) model  \cite{RN101,RN72,RN99}, a supernova could provide large anti-neutrino fluxes to induce molecular homochirality in a close meteoric environment through inverse beta decay interactions, which favors one of the enantiomers.  Although this mechanism always produces the same sign of the chirality,  it can generate only a small amount of enantiomeric excess (around $10^{-6}$ ) \cite{RN72}.

In some studies, two major paradigms autocatalytic processes and parity-violating interactions, are combined by adding a term of parity violation in the equation of state of the system \cite{Kondepudi1985,RN74,RN71}.  The noise-induced system in far-from-equilibrium in the absence of  parity-violation effects has been considered  previously \cite{RN70}.

In this paper, parity-violating interaction as a direct term in the chemical scheme is included, where we take  a noise-induced system in a far-from-equilibrium condition, where parity-violating effects are considered as a direct term in the chemical scheme. Furthermore, we demonstrate that neutrino-chiral molecules should be incorporated as a deterministic part in our dynamical equations and chemical scheme, rather than being treated as noise or a bias contribution. We will assume an IMC close to a supernova event, where autocatalytic processes could occur at the same time that a large flux of supernova neutrinos triggered PVED between two enantiomeric states. By creating unequal activation energies for the forward (L → D) and reverse (D → L) reactions, PVED establishes a directional bias in the racemization process. 
We take into account fluctuations in the system and consider a far-from-equilibrium condition to be more realistic. In our scenario, a meteorite could finally carry  chiral molecules to Earth as the origin of life. Our mechanism combines both major paradigms regarding the origin of homochirality; additionally, this mechanism includes stochastic processes in  a far-from-equilibrium condition. Meanwhile, this model explains a possible answer for the enantiomeric excess that exists on the mentioned meteorites.

In section \ref{II}, we will discuss the interaction of neutrinos with chiral molecules and the PVED.
In section \ref{Chem}, we will discuss the chemical scheme of our model and how we apply the PVED into the model.  
In section \ref{IV}, we will find the stochastic equations of the system. 
In section \ref{V}, the dynamics of the probability distribution of the system taking into account a reliable physical parameters will be presented, and finally, in the section \ref{VI}, we will conclude our discussion.

\section{\label{II} Supernova Neutrino Interaction with Chiral Molecules}
In supernova events, an intense flux of neutrinos is generated, which gradually diminishes over time \cite{giunti2007fundamentals}. Additionally, the distance from the supernova event is crucial as its emitted radiation could destroy complex molecules such as amino acids or other chiral molecules. The estimated distance from which this kind of molecule survives supernova radiation is 0.01 AU (Astronomical Unit) \cite{reach2024supernova, famiano2018selection}.
Instead of using a time variable function for neutrino flux, we use the average luminosity of such events around the peak of their luminosity, which is on the order of \(10^{51}-10^{53}erg/s\) and the average energy of electron neutrinos, which is \(10 \) MeV \cite{giunti2007fundamentals}. %Due to the duration of the peak of the luminosity in supernova events, in the best case, this simplification is correct for a timescale of few days and no longer.
Using these values, flux of the electron neutrinos in 0.01 AU distance from the supernova core is in the order of order of \(10^{33}-10^{35} cm^{-2}s^{-1}\) which is equal to a number density of \(10^{23}-10^{25} cm^{-3}\) or around \(10^3-10^5\) M (Molar).

Considering only electron neutrinos, one could find the effective Lagrangian for neutrino-electron interaction as \cite{RN37}:
\begin{align}
\mathcal{L}_{\text{eff}}
&= -\sqrt{2}\, G_F\,
\big( \bar{u}_e \gamma_{\mu} (g_L P_L + g_R P_R) u_e \big) \times \notag \\
&\quad \big( \bar{u}_{\nu} \gamma^{\mu} P_L u_{\nu} \big),
\end{align}
where 
$P_{R,L} = \tfrac{1}{2}(1 \pm \gamma^5)$, 
$g_L = 2\sin^2\theta_W + 1$, 
$g_R = 2\sin^2\theta_W$, 
in which $\theta_W$ is the weak mixing angle and $\sin^2\theta_W = 0.23$.
Here, $u_{\alpha}$ denotes the spinor of the particle $\alpha$, $\bar{u}_{\alpha}$ its Dirac adjoint, 
and $G_F = 1.166 \times 10^{-5}\,\mathrm{GeV}^{-2}$ is the Fermi coupling constant. The Hamiltonian density of this interaction is:
\begin{align}
\label{H}
H_{\nu-e} 
 & = \frac{\sqrt{2}}{2} G_F\, (g_L P_L + g_R P_R)\, n_{\nu}& \\ \nonumber
& = \frac{G_F}{2\sqrt{2}} (4 sin^2 \theta_W +1 - \gamma ^5 ) n_\nu .
\end{align}

Here, we assume the antisymmetric flux of the Dirac neutrino and antineutrino where \(n_\nu - n_{\bar{\nu}} \approx n_\nu \). This could be considered as a perturbation Hamiltonian, in which the \(\gamma ^{5}\) term causes separation in the eigen energies of the enantiomeric molecular states \cite{RN25}. The spin-orbit corrections are necessary because the expectation value of \(\gamma ^{5}\) vanishes for pure singlet and triplet states, and the spin-orbit effect mixes these states \cite{RN27}. %The order of PVED has been studied in \cite{RN27}:

One could find the PVED using the difference between expectation values of the perturbation Hamiltonian for opposite enantiomeric molecular states as:
\begin{eqnarray}
\Delta E_{\mathrm{PV}} \sim {G_F}  \, \Delta_{R,S} \langle \gamma^5 \rangle \, n_{\nu}.
\end{eqnarray}

In \cite{RN27}, the order of \(\Delta_{R,S} \langle \gamma^5 \rangle\) for CHBrClF as a typical chiral molecule was estimated using the functional density theory with functional density B3LYP, \(\Delta_{R,S} \langle \gamma^5 \rangle\approx10^{-10}\). Assuming the validity of these approximations for amino acids, one could find the  estimated value for PVED  as follows:
\begin{eqnarray}\label{e4}
\Delta E_{\mathrm{PV}} \sim 10^{-22} \left( \frac{n_{\nu}}{10^{25}\text{cm}^{-3}} \right)~\text{eV}.
\end{eqnarray}

Here, the source of PVED is an external neutrino-induced interaction and   is proportional to  the neutrino flux. While in the previous studies  \cite{Kondepudi1985,RN74,RN71,RN88,RN86,RN85}, internal weak interactions break  the chiral symmetry  between two enantiomers.

\section{\label{Chem}{Chemical Scheme}}
In this work, our goal is to emphasize the role of supernova neutrino flux on the emergence of biomolecular homochirality. To do so, we study a chemically closed system of achiral molecules and left-handed and right-handed chiral molecules, which chemically interact with each other, in the presence of the PVED induced by neutrinos. There are some studies \cite{RN88,RN85,RN86} in which an open system was considered, including input and output flows of molecules. In contrast to \cite{RN70}, we also consider the microscopic reversibility, to present a more general framework. Following the experimental result \cite{sczepanski2014cross}, the chiral inhibition reaction is not a necessary part of the chemical scheme in the auto-catalytic scenario, so we do not consider this type of reaction. 
 
\subsection{Autocatalytic and Non-autocatalytic Reactions}
In our model, we considered the autocatalytic reactions for left-handed molecules (L) and right-handed molecules (D), as follows:
\begin{eqnarray}
\mathrm{A} + \mathrm{D} 
\underset{k_{-a}}{\overset{k_a}{\rightleftharpoons}} 
\mathrm{D} + \mathrm{D},
\label{eq:reaction}
\end{eqnarray}
\begin{eqnarray}
\mathrm{A} + \mathrm{L} 
\underset{k_{-a}}{\overset{k_a}{\rightleftharpoons}} 
\mathrm{L} + \mathrm{L},
\label{eq:reaction6}
\end{eqnarray}
where $k_a$ is the rate constant of the forward autocatalytic reaction, $k_{-a}$ is the rate constant of the backward reaction, and \textit{A} stands for achiral molecules.
Because the activation energies of different enantiomers are close, we considered the same rates for these reactions.

In the chemical scheme, there are decay reactions where chiral molecules convert to achiral ones; due to detailed balance, their backward reactions, known as non-autocatalytic decay, must be considered too:
\begin{eqnarray}
\mathrm{D} 
\underset{k_{-d}}{\overset{k_{d}}{\rightleftharpoons}} 
\mathrm{A},
\label{eq:reactionee}
\end{eqnarray}
\begin{eqnarray}
\mathrm{L} 
\underset{k_{-d}}{\overset{k_{d}}{\rightleftharpoons}} 
\mathrm{A},
\label{eq:reactiondirect}
\end{eqnarray}
where $k_d$ and $k_{-d}$ indicate the rate constants of chiral decay and non-autocatalytic reactions, respectively.
This chemical scheme might be compared to the work of Jafarpour \cite{RN70}, but it additionally considers the backward reactions of the autocatalytic reactions.
\subsection{Parity Violating Interaction in the Chemical Scheme}
The role of parity-violating interactions in the origin of biomolecular homochirality has been studied in \cite{RN1,RN17,RN18,RN19,RN16,RN3,RN46,RN45}. Although some works \cite{RN88,RN85,RN86,RN74,RN71,Kondepudi1985} tried to consider this effect in the autocatalytic scenario, none of them applied it for external weak interactions, where the neutrino flux could play a role in the dynamics of the system.

\begin{figure}[!htbp]
\includegraphics[width=0.4\textwidth]{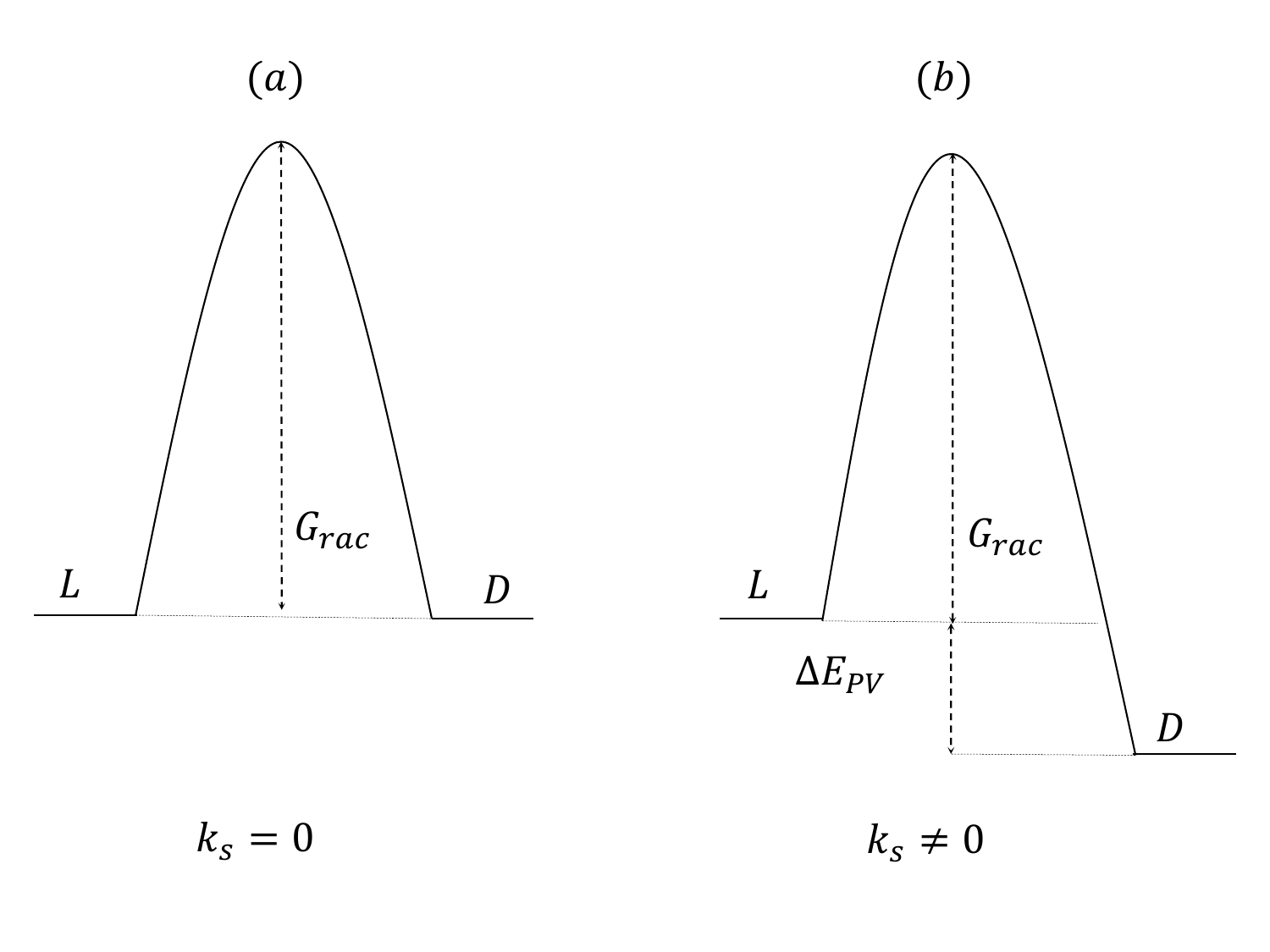}% Here is how to import EPS art
\caption{\label{raceee} Activation energy of the racemization reaction (a) without and (b) with applying parity-violation interaction.} 
\end{figure}

A large number of neutrinos generated by a nearby supernova explosion could enter the system, and consequently, a fraction of them interact with molecules. Because of the weak interactions, neutrinos interact differently with left-handed and right-handed chiral molecules and induce a small difference in their activation energy for racemization reactions. This energy difference is PVED, which is shown in Fig. \ref{raceee}. Note that the favored handedness is not universal; the sign and magnitude of the PVED depend sensitively on the specific molecular structure. Here, we assumed that the small PVED is in favor of the stability of the right-handed molecules. When there is no PVED, the L(D)-enantiomer needs the same activation energy of $G_{rac}$ for converting to the D(L)-enantiomer, but in the presence of PVED, the D-enantiomer needs to take the extra $\Delta E_{PV}$ to convert to the opposite enantiomer, while the L-enantiomer needs the same activation energy.

\begin{figure}[!htbp]
\includegraphics[width=0.4\textwidth]{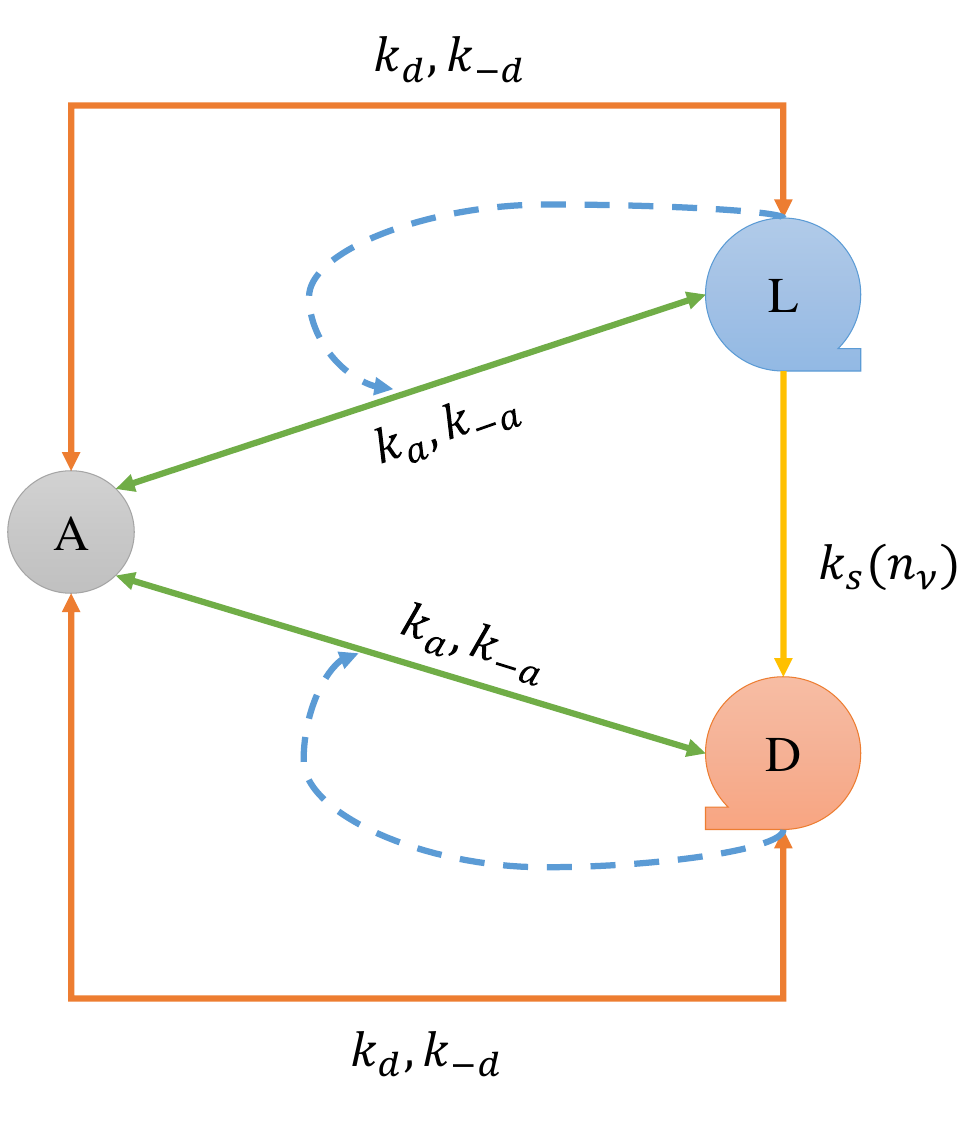}% Here is how to import EPS art
\caption{\label{ChS} Chemical scheme of the system, Eqs. (\ref{eq:reaction}) - (\ref{eq.6}), considering enantiomer-neutrino interactions. The green arrows represent the autocatalytic reactions, the red arrows represent the non-autocatalytic decays and their backward reactions, and finally, the yellow arrow represents the neutrino-enantiomer reaction.}
\end{figure}

The PVED could be considered as a chemical reaction:
\begin{eqnarray} 
\ L \xrightarrow{k_s} D,
\label{eq.6}
\end{eqnarray}

where $k_s$ is the effective rate constant of this reaction, which is a function of neutrino flux. In general, PVED introduces a slight energetic preference between enantiomers. 
Since the activation energy is not too large, one might consider the following reaction:
\begin{eqnarray} 
\mathrm{L} 
\underset{k_{-rac}}{\overset{k_{rac}}{\rightleftharpoons}} 
\mathrm{D}.
\end{eqnarray}
 Note that in the racemic limit, the effective rate constant $k_s \approx k_{rac}-k_{-rac}$ is a good approximation. In fact, applying reaction  (\ref{eq.6}) means that the reverse reaction is also effectively included. The chemical scheme is shown in Fig. \ref{ChS}.


Due to transition state theory \cite{wright1999fundamental,kondepudi2014modern, RN86}:
%One could calculate the order of  $k_s$ using 
\begin{equation}
\label{rate constant}
k(T) = \frac{k_B T}{h} \, \exp\left(-\frac{\Delta G^\ddagger}{RT}\right) 
\left( \mathrm{M}^{1-m} \, \mathrm{s}^{-1} \right),
\end{equation}
where $k_B$ is the Boltzmann constant, $T$ is the temperature, $h$ is the Planck constant, 
$\Delta G^\ddagger$ is the Gibbs free energy of activation per mole, for reactions with one reactant $m=1$, and for two reactants $m=2$, 
and $R=N_{A} k_{B}$ is the gas constant. 

Using this formula and considering the difference between the racemization reactions due to parity-violation interactions as the $k_s$ reaction % favoring D enantiomer.
%The chemical scheme presented in Sec. \ref{Chem} might be extended to include. 
and replacing the Gibbs free energy of activation per mole with the energy of activation per mole as a simplification (using the Arrhenius equation \cite{kondepudi2014modern}), one could find:
\begin{align}  
\label{k_s}
k_s =& \frac{k_B T}{h} \,
\exp\left(-\frac{G_{rac}}{R T}\right) \\ \nonumber
&\times\left( \exp\left(\frac{\Delta E_{\mathrm{PV}}}{k_B T}\right) - 1 \right) (s^{-1}),
\end{align}
where \(G_{rac}\) is the activation energy of the racemization reaction per mole.
As PVED is too small, one could approximate this equation as:
\begin{equation}
\label{k_S}
k_s \approx \frac{\Delta E_{\mathrm{PV}}}{h} \,
\exp\left(-\frac{G_{rac}}{R T}\right).
\end{equation}

There are different computational estimations and experimental results for the amino acids' racemization activation energies.  For example, In \cite{aubrey2008amino}, for an astronomical environment, this parameter has been calculated for several amino acids: glycine (28.1 kJ/mol), alanine (39.9 kJ/mol), valine (26.3 kJ/mol), glutamic acid (37.9 kJ/mol), and serine (49.3 kJ/mol) with an Arrhenius pre-exponential factor of the same order of transitions state theory. We estimate the order of $k_s$ using the following values:

\begin{align}
k_s &\sim 10^{-16}\mathrm{\,s^{-1}}
\left( \frac{\Delta E_{\mathrm{PV}}}{10^{-22}\ \mathrm{eV}} \right)
 \nonumber \\
&\quad \times
\exp \Bigg[
-\left( \frac{G_{rac}}{26.3\ \mathrm{kJ/mol}} \right)
\left( \frac{165\ \mathrm{K}}{T} \right)
\Bigg].
\end{align}

Assuming $G_{rac}\sim  26.3 \ \mathrm{kJ/mol}$ and $\Delta E_{PV} \sim 10^{-22} \ \mathrm{eV}$, for $130 \ \mathrm{K}<T<205 \ \mathrm{K}$, $k_s$ changes from $6.5\times10^{-19} s^{-1}$ to $4.8\times 10^{-15} s^{-1}$ (see Table. \ref{tab:racemization} and Table \ref{tab:high_temp_rates}).

\section{\label{IV}{Stochastic Equations of the System}}
To study a system in far-from-equilibrium conditions, we find transition rates, master, Fokker-Planck, and Langevin equations, and we consider the evolution of the system in the Itô sense following the approach presented in \cite{RN70}.
\subsection{Stochastic Equation in Itô Sense}
\subsubsection{Master Equation}

The state of the system can be described by the following state vector:
\begin{equation}
x = \begin{bmatrix}
a \\
d \\
l
\end{bmatrix},
\end{equation}
where \textit{a}, \textit{d}, and \textit{l} are the concentrations of \textit{A}, \textit{D}, and \textit{L} molecules, respectively.  

The chemical scheme, Eqs. (\ref{eq:reaction}) - (\ref{eq.6}), could be represented by the \textit{stoichiometry matrix} as below:
%One could introduce a matrix which represents the chemical scheme. This matrix is called the \textit{stoichiometry matrix}:
\begin{equation}
S = \begin{bmatrix}
-1 & +1 & 0 \\
-1 & 0 & +1 \\
+1 & -1 & 0 \\
+1 & 0 & -1 \\
0 & +1 & -1
\end{bmatrix}.
\label{Stio}
\end{equation}
Each row of this matrix shows chemical reactions in which a molecule is consumed, and another is produced.

We determine the transition rates for the state of the system, using the chemical kinetics for the reactions Eqs. (\ref{eq:reaction}) - (\ref{eq.6}) as follows:

\begin{align}
T(\mathbf{x} + \varepsilon \mathbf{s}_1 \mid \mathbf{x}) &= k_a a d + k_{-d} a,
\label{21}\\
T(\mathbf{x} + \varepsilon \mathbf{s}_2 \mid \mathbf{x}) &= k_a a l + k_{-d} a, \label{22}\\
T(\mathbf{x} + \varepsilon \mathbf{s}_3 \mid \mathbf{x}) &= k_{-a} d^2 + k_d d, \label{23}\\
T(\mathbf{x} + \varepsilon \mathbf{s}_4 \mid \mathbf{x}) &= k_{-a} l^2 + k_d l,\label{24}\\
T(\mathbf{x} + \varepsilon \mathbf{s}_5 \mid \mathbf{x}) &= k_s l,\label{25}
\end{align}
where $\varepsilon=(N_{A}V)^{-1}$ of the system, and \textbf{\(s_i\)} (ith row of the Eq. (\ref{Stio})) represents the direction of each stoichiometry vector.

\begin{figure}[!htbp]
\includegraphics[width=0.4\textwidth]{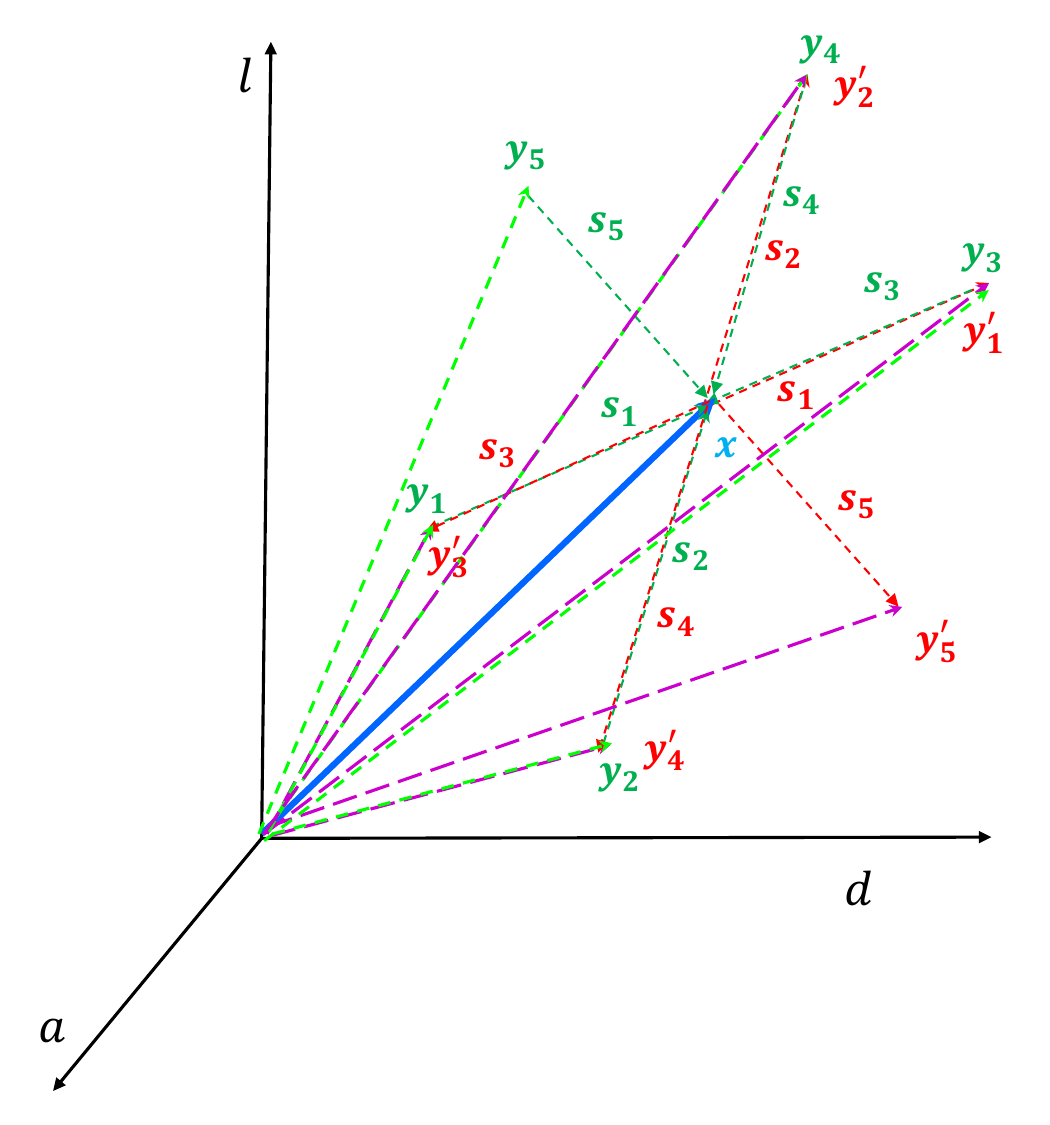}% Here is how to import EPS art
\caption{\label{fig.5} The schematic picture of how Eq~(\ref{eq.21}) describes the time evolution of \(P(x,t)\) due to the stoichiometry matrix.}
\end{figure}

The time evolution of the probability of the system at a given state \textbf{x} is described through the following master equation:

\begin{eqnarray}
\frac{\partial P(\mathbf{x}, t)}{\partial t} 
& = & 
V (\sum_{\mathbf{y}} T(\mathbf{x} \mid \mathbf{y}) P(\mathbf{y}, t) \nonumber \\
& & 
-  \sum_{\mathbf{y}} T(\mathbf{y} \mid \mathbf{x}) P(\mathbf{x}, t) ).
\label{eq.20}
\end{eqnarray}
 In our system, Eq. (\ref {eq.20}) turns into:
 %we are studying, the next and previous states are given by stoichiometry vectors, and due to transition rates, one could find:
\begin{eqnarray}
\frac{\partial P(\mathbf{x}, t)}{\partial t} &=& 
V ( \sum_{i=1}^{5} T(\mathbf{x} \mid \mathbf{x} - \varepsilon \mathbf{s}_i) P(\mathbf{x} - \varepsilon \mathbf{s}_i, t) \nonumber \\
&& - \sum_{i=1}^{5} T(\mathbf{x} + \varepsilon \mathbf{s}_i \mid \mathbf{x}) P(\mathbf{x}, t)).
\label{eq.21}
\end{eqnarray}
Fig~\ref{fig.5} shows the schematic picture of the above master equation.
%could help to understand how this equation is found.

\subsubsection{Fokker-Planck Equation}
%To find Fokker-Planck equation in the Itô Sense, it is convinient to define following function \cite{RN70}:

We obtain the Fokker-Planck equation in the Itô sense as \cite{RN70}: 

\begin{eqnarray}
\frac{\partial P(\mathbf{x}, t)}{\partial t} & \approx & 
- \sum_{j=1}^3 \frac{\partial}{\partial x_j} \left( H_j \, P(\mathbf{x}, t) \right) \nonumber \\
& & + \frac{1}{2} \sum_{j=1}^3 \sum_{k=1}^3 \frac{\partial^2}{\partial x_j \partial x_k} \left( B_{jk} \, P(\mathbf{x}, t) \right),
\label{eq.26}
\end{eqnarray}
where $\textbf{H}$ is the drift vector:
\begin{align}
\mathbf{H} = \sum_{i=1}^{5} T(\mathbf{x} + \varepsilon \mathbf{s}_i \mid \mathbf{x}) \, \mathbf{s}_i,
\label{eq.27}
\end{align}
and $B$ is the diffusion matrix:
\begin{align}
B_{jk} = \varepsilon \sum_{i=1}^{5} S_{ij} S_{ik} \, T(\mathbf{x} + \varepsilon \mathbf{s}_i \mid \mathbf{x}).
\label{eq:B_jk}
\end{align}

In our system, one could find:
\begin{align}
{\mathbf{H}} = 
\begin{bmatrix}
- k_a a r - 2 k_{-d} a + k_{-a} (d^2 + l^2) + k_d r \\
k_a a d + k_{-d} a - k_{-a} d^2 - k_d d + k_s l \\
k_a a l + k_{-d} a - k_{-a} l^2 - k_d l - k_s l
\end{bmatrix},
\label{eq.28}
\end{align}
and:
\begin{widetext}
{\small
\begin{equation}
\frac{B}{\varepsilon} =
\begin{bmatrix}
(k_a a + k_d)r + 2 k_{-d} a + k_{-a} (d^2 + l^2)&-(k_a a +k_d+k_{-a} d) d - k_{-d} a& -(k_a a +k_d+k_{-a} l) l - k_{-d} a \\
-(k_a a +k_d+k_{-a} d) d- k_{-d} a&(k_a a +k_d+k_{-a} d) d + k_{-d} a + k_s l&- k_s l \\
- (k_a a +k_d+k_{-a} l) l - k_{-d} a&- k_s l&(k_a a +k_d+ k_{-a} l+ k_s) l + k_{-d} a 
\end{bmatrix},
\label{eq.29}
\end{equation}}
\end{widetext}
where $r=d+l$ is the total concentration of chiral molecules.

Note that, in the continuous limit where we have a large number of molecules, the Fokker-Planck equation follows the second-order approximation independent of \(\varepsilon\) value.

%Then we found the Fokker-Planck equation in Itô sense for the system.

\subsubsection{Langevin Equation}
The stochastic dynamics of the system is described by the Langevin equation as \cite{RN127}:
%One could find the Langevin equation of the system as follows :
\begin{equation}
\frac{d \mathbf{x}}{dt} = \mathbf{H}(\mathbf{x}) + \boldsymbol{\xi}(t),
\label{eq.31}
\end{equation}
where $\xi$ is a zero-mean Gaussian noise vector with the following autocorrelation function:
\begin{equation}
\langle \xi_i(t), \xi_j(t') \rangle = B_{ij} \, \delta(t - t').
\label{eq.32}
\end{equation}

It is shown in \cite{RN70} that one could rewrite the Langevin equation in terms of Gaussian white noise by decomposing the diffusion matrix. By defining the $G$ matrix, we can decompose the B matrix as:
%Thus, we define $G$ as follows:
\begin{equation}
B = G G^{T}.
\end{equation}

To find $G$, we use the following relation as discussed in the appendix of \cite{RN70}:
\begin{equation}
G_{ik} = \sqrt{\varepsilon} \, \sqrt{T(\mathbf{x} + \varepsilon \mathbf{s}_k \mid \mathbf{x})} \, S_{k i}.
\label{eq.34}
\end{equation}

For convenience, one could present the reduced form of the stoichiometry matrix (Eq. (\ref{Stio})) as follows:
\begin{equation}
S^r =
\begin{bmatrix}
-1 & +1 & 0 \\
-1 & 0 & +1 \\
0 & +1 & -1
\end{bmatrix},
\label{eq.35}
\end{equation}
and the corresponding transition rates are:
\begin{eqnarray}
T_1^r &=& k_a a d + k_{-d} a + k_{-a} d^2 + k_d d, \\
T_2^r &=& k_a a l + k_{-d} a + k_{-a} l^2 + k_d l, \\
T_3^r &=& k_s l.
\end{eqnarray}

Using these parameters, we find:
\begin{widetext}
\begin{align}
G = \sqrt{\varepsilon}
\begin{bmatrix}
-\sqrt{k_a a d + k_{-d} a + k_{-a} d^2 + k_d d} & -\sqrt{k_a a l + k_{-d} a + k_{-a} l^2 + k_d l} & 0 \\
 \sqrt{k_a a d + k_{-d} a + k_{-a} d^2 + k_d d} & 0 &\sqrt{k_s l}  \\
0 & \sqrt{k_a a l + k_{-d} a + k_{-a} l^2 + k_d l} & -\sqrt{k_s l}
\end{bmatrix},
\label{eq.36}
\end{align}
\end{widetext}
and the Langevin equation in terms of white noise $\eta$ is given by:
\begin{equation}
\frac{d\mathbf{x}}{dt} = \mathbf{H}(\mathbf{x}) + \mathbf{G} \, \boldsymbol{\eta}(t).
\label{eq.37}
\end{equation}
%where its autocorrelation function is:
%\begin{equation}
%\langle \eta_i(t), \eta_j(t') \rangle = \delta_{ij} \, \delta(t - t')
%\label{eq.38}
%\end{equation}
Note that the dynamical evolution of \textbf{x} is obtained by solving this equation, which consequently determines the probability distribution of the system.

\subsection{Stochastic Equations of Enantiomeric Excess}
In order to study the enantiomeric excess of the system $( \omega= \frac{[D]-[L]}{[D]+[L]})$, we rewrite our equations in terms of a state vector including $\omega$. 
Here, we consider a closed system in which the deterministic part of the total concentration of the molecules (n) is :
\begin{equation}
\frac{d n}{dt} = \frac{d}{dt}(a + d + l) = 0.
\label{eq.47}
\end{equation}

Note that the total concentration of molecules might be changed due to stochastic fluctuations.
In fact, there is a Gaussian distribution around a certain value $n^*$.
Moreover, the deterministic term of time evolution of the total chiral molecules concentration ($r$) is:
\begin{equation}
\begin{split}
\dot{r} = & -\left(k_a + \tfrac{1}{2} k_{-a} (\omega^2 + 1)\right) r^2 \\
& + \left(k_a n - 2k_{-d} - k_d\right) r + 2k_{-d} n
\label{eq.rdot}.
\end{split}
\end{equation}
Taking into account the value of the parameters in the scenario presented in Sec. \ref{V}, $\dot r\approx 0$ can be used as a reliable  approximation. Therefore, the deterministic part of \textit{r} has almost no dynamics, and we can use the following constraint to reduce the degrees of freedom:
\begin{equation}
\frac{d r}{dt} = \frac{d}{dt}(d + l) \approx 0.
\label{eq.48}
\end{equation}
By solving the above equation, it is obtained that r has  a distribution around a certain value $r^*$ given by:
\begin{widetext}
\begin{eqnarray}
r^{*} = \frac{
    k_a n - 2 k_{-d} - k_d
}{%
    2 \left( k_a + \frac{1}{2} k_{-a} (\omega^2 + 1) \right)
} 
 + \frac{
    \sqrt{
        \left( k_a n - 2 k_{-d} - k_d \right)^2 + 8 k_{-d} k_a n + 4 k_{-d} k_{-a} n (1 + \omega^2)
    }
}{%
    2 \left( k_a + \frac{1}{2} k_{-a} (\omega^2 + 1) \right)
}.
\label{eq.49}
\end{eqnarray}
\end{widetext}

We see that $r^*$ depends on time through enantiomeric excess ($\omega$). However, as we will see, $k_{-a}$ is negligible in comparison with $k_{a}$, in our desired parameter space. Therefore, we can neglect $k_{-a}$ terms which also includes $\omega$ in favour of $k_{a}$ leading to following equation:
\begin{eqnarray}
r^{*} &\approx& \frac{
    k_a n - 2 k_{-d} - k_d
}{2 k_a} \nonumber \\
&& + \frac{
    \sqrt{
        \left( k_a n - 2 k_{-d} - k_d \right)^2 + 8 k_{-d} k_a n
    }
}{2 k_a}.
\label{eq.51}
\end{eqnarray}
We introduce \textbf{y} state vector as below in order to obtain stochastic differential equations of $\omega$:
%Using this approximation, constraints, and the importance of enantiomeric excess, the following state vector could be the best choice that we could make:
\begin{equation}
\mathbf{y} =
\begin{bmatrix}
n \\
r \\
\omega
\end{bmatrix}
\equiv
\begin{bmatrix}
a + d + l \\
d + l \\
(d - l)/(d + l)
\end{bmatrix}.
\label{eq.53}
\end{equation}

Taking into account Eqs. (\ref{eq.47}) and (\ref{eq.48}), the dynamics of this new state vector are determined by $\dot{\omega}$. As we now have only one variable, solving stochastic equations is straightforward. $\dot{\omega}$ can be evaluated through Eq. (\ref{eq.37}) and using the rule of linear combination of Gaussian variables:
{\small \begin{eqnarray}
\dot{\omega} &=&  \Bigg[
    \frac{1}{2} k_{-a} \, r^{*}(\omega^{3} - \omega)
    + \left(k_{s} (1 - \omega) + 2 k_{-d} \omega \right)
    - \frac{2 k_{-d} N}{Vr^*} \, \omega
\Bigg] \nonumber \\
&& + \sqrt{
    \frac{\omega^{2} + 1}{N_AV} \left[
        \frac{2 k_{d}+ k_{s} (1 - \omega)}{r^{*}}  + k_{-a} (\omega^{2} + 1)
    \right]
} \, \eta(t),
\label{eq:omega_sde_broken}
\end{eqnarray}}where N is the total number of molecules originating from the relation $n=N/V$. 
Regarding Eqs.  (\ref{eq:B_jk}) and (\ref{eq:omega_sde_broken}), the Fokker-Planck equation of the system is:
\begin{widetext}
\begin{eqnarray}
\frac{\partial P(\omega, t)}{\partial t} 
& \simeq & 
- \frac{\partial}{\partial \omega} \left(
\frac{1}{r^{*}} \left[
\frac{1}{2} k_{-a} \, \omega (\omega^{2} - 1) (r^{*})^{2} 
+ \left( k_{s} (1 - \omega) + 2 k_{-d} \omega \right) r^{*} 
- \frac{2 k_{-d} N}{V} \omega 
\right] P(\omega, t)
\right) \nonumber \\
&& + \frac{1}{2} \frac{\partial^{2}}{\partial \omega^{2}} \left(
\frac{\omega^{2} + 1}{N_A V r^{*}} \left[
2 k_{d} + k_{s} (1 - \omega) + k_{-a} (\omega^{2} + 1) r^{*}
\right] P(\omega, t)
\right).
\label{eq:FP_omega_broken}
\end{eqnarray}

\end{widetext}

We solve this equation in the case of estimated parameters of an IMC in the environment around a supernova explosion.
\section{\label{V}{Results}}
%\subsection{Estimation of Model Parameters}
 %\textcolor{red}{refrences required:}
IMCs reach a size of tens of parsecs, and their typical temperature is around 10 K to 100 K \cite{bally1986interstellar}; however, when a special event such as a supernova occurs, the temperature of a nearby IMC could be increased up to 3000 K \cite{kortgen2016supernova}. %text however, as the distance increases the temperature falls. 
The chemical composition of IMCs contains various types of simple and complex molecules ranging from $H$, $H_2$ to complex organic molecules such as $CH_3OH$, $CH_3 OCH_3$, $CH_3OCHO$ and $CH_3CHOCH_2$  \cite{Nomura:2003fm}.  In \cite{belloche2009increased}, it was shown that the column density of complex molecules such as  dimethyl ether (\(CH_3 OCH_3\)), in the IMCs should be \(n_c\leq10^{18}\   \text{cm}^{-2}\). The number density of molecules can be obtained from $n=n_c/L$ where $L$ is the thickness of the IMC and is assumed to be a few 0.01 pc \cite{andre2010filamentary,Nomura:2003fm}.

In our numerical calculations, we consider the total density of the molecules in the chemical scheme \(n\approx 100 \ \text{cm}^{-3}\approx 10^{-18}M\). Also, as an approximation, we take the IMC as a homogeneous environment and the system as a part of it. The Volume of the system can be estimated by the swept volume by the meteorite; the typical size of meteorites is in the order of several meters, so due to the depth of the IMC ($10^{-2}$pc), 
we consider the swept volume in the range of $10^{5} km^3$ to $10^{7} km^3$\cite{rubin2010meteorite}. As the supernova occurs near the IMC, the temperature could vary from 10 K to 3000 K. In Table. \ref{tab:racemization} and Table. \ref{tab:high_temp_rates}, we calculate the rate constants of the chemical reactions at different temperatures using Eq. (\ref{rate constant}). $G_{rac}$ also presents the value of \(k_s\) for different temperatures using Eq. (\ref{k_s}), for a supernova with a typical luminosity of \(10^{53}\) erg/s, and an IMC located at a distance of 0.01 AU from it.

Regarding Table. \ref{tab:racemization} and Table \ref{tab:high_temp_rates}, although chemical equilibrium between forward and backward chemical reactions is held, at low temperatures, the backward rate constant  is suppressed. In fact, the backward reactions take longer than the age of the universe to occur. Although in low temperatures, microscopic reversibility  is allowed \cite{blackmond2009if}, there will be no observational evidence on this principle. According to \cite{cline1996homochiral,cline2005physical}

%\begin{widetext}
\begin{table*}[htbp]
\centering
\caption{Rate constants $k$ and activation free energies $\Delta G^\ddagger$ for different temperatures.}
\label{tab:racemization}

\renewcommand{\arraystretch}{1.4} % Increase row height
\setlength{\tabcolsep}{6pt}      % Increase column spacing

\begin{tabular}{lccccccc}
\hline\hline
Reaction &
$\Delta G^{\ddagger}$ [kJ mol$^{-1}$] &
130 K &
150 K &
160 K &
165 K &
170 K &
175 K \\
\hline

$k_a$ [M$^{-1}$ s$^{-1}$] &
62.15\cite{RN86}  &
$2.8\times10^{-13}$ &
$7.0\times10^{-10}$ &
$1.7\times10^{-8}$ &
$7.1\times10^{-8}$ &
$2.8\times10^{-7}$ &
$1.0\times10^{-6}$ \\

$k_{-a}$ [M$^{-1}$ s$^{-1}$] &
90.93 \cite{RN86}  &
$7.6\times10^{-25}$ &
$6.5\times10^{-20}$ &
$6.7\times10^{-18}$ &
$5.5\times10^{-17}$ &
$4.0\times10^{-16}$ &
$2.6\times10^{-15}$ \\

$k_d$ [s$^{-1}$] &
96.68 \cite{RN86}  &
$3.7\times10^{-27}$ &
$6.5\times10^{-22}$ &
$8.8\times10^{-20}$ &
$8.3\times10^{-19}$ &
$6.8\times10^{-18}$ &
$4.9\times10^{-17}$ \\

$k_{-d}$ [s$^{-1}$] &
125.47 \cite{RN86} &
$9.9\times10^{-39}$ &
$6.1\times10^{-32}$ &
$3.5\times10^{-29}$ &
$6.3\times10^{-28}$ &
$1.0\times10^{-26}$ &
$1.2\times10^{-25}$ \\

$k_s$ [s$^{-1}$] &
26.30 \cite{aubrey2008amino}&
$6.5\times10^{-19}$ &
$1.7\times10^{-17}$ &
$6.2\times10^{-17}$ &
$1.1\times10^{-16}$ &
$2.0\times10^{-16}$ &
$3.4\times10^{-16}$ \\
\hline\hline
\end{tabular}
\end{table*}
\begin{table*}[htbp]
\centering
\caption{Rate constants $k$ and activation free energies $\Delta G^\ddagger$ for different temperatures.}
\label{tab:high_temp_rates}

\renewcommand{\arraystretch}{1.4} 
\setlength{\tabcolsep}{6pt}      

\begin{tabular}{lccccccc}
\hline\hline
Reaction &
$\Delta G^{\ddagger}$ [kJ mol$^{-1}$] &
180 K &
185 K &
190 K &
195 K &
200 K &
205 K \\
\hline

$k_a$ [M$^{-1}$ s$^{-1}$] &
62.15 \cite{RN86} &
$3.4\times10^{-6}$ &
$1.1\times10^{-5}$ &
$3.2\times10^{-5}$ &
$9.0\times10^{-5}$ &
$2.4\times10^{-4}$ &
$6.1\times10^{-4}$ \\

$k_{-a}$ [M$^{-1}$ s$^{-1}$] &
90.93 \cite{RN86} &
$1.5\times10^{-14}$ &
$8.0\times10^{-14}$ &
$3.9\times10^{-13}$ &
$1.7\times10^{-12}$ &
$7.3\times10^{-12}$ &
$2.8\times10^{-11}$ \\

$k_d$ [s$^{-1}$] &
96.68 \cite{RN86} &
$3.2\times10^{-16}$ &
$1.9\times10^{-15}$ &
$1.0\times10^{-14}$ &
$5.0\times10^{-14}$ &
$2.3\times10^{-13}$ &
$9.7\times10^{-13}$ \\

$k_{-d}$ [s$^{-1}$] &
125.47 \cite{RN86} &
$1.4\times10^{-24}$ &
$1.4\times10^{-23}$ &
$1.2\times10^{-22}$ &
$9.6\times10^{-22}$ &
$6.9\times10^{-21}$ &
$4.4\times10^{-15}$ \\

$k_s$ [s$^{-1}$] &
26.30 \cite{aubrey2008amino}&
$5.6\times10^{-16}$ &
$9.0\times10^{-16}$ &
$1.4\times10^{-15}$ &
$2.2\times10^{-15}$ &
$3.2\times10^{-15}$ &
$4.8\times10^{-15}$ \\
\hline\hline
\end{tabular}
\end{table*}

Before discussing the results, we would like to justify the constraint (\ref{eq.48}) as a good approximation. To do so, we will use our parameter space to find the order of magnitude of terms in Eq. (\ref{eq.rdot}). For T=165 K, \(k_a\approx 10^{-7} M^{-1}s^{-1}\), \(k_{-a}\approx 10^{-17}M^{-1}s^{-1}\) and \(-1\leq \omega \leq +1\), then \(-k_a r^2\) is dominant in the first term of the Eq. (\ref{eq.rdot}). For the second and third terms, we have \(k_a n\approx 10^{-25}s^{-1}\), \(k_d\approx 10^{-18} s^{-1}\) and \(k_{-d} \approx 10^{-27} s^{-1}\); therefore the dominant term is \(-k_d r\). The total concentration of the chiral molecules cannot be more than the total concentration of the molecules; so \(r\leq n\approx 10^{-18}M\). 
So  Eq. (\ref{eq.rdot}) can be approximated as:
\begin{equation}
\frac{d r}{dt} \approx -k_d r  \rightarrow  r(t)\approx r_0 \text{exp}(-k_d t).
\label{eq.approx rdot}
\end{equation}

Noticing the negligible value of $k_{d}$ for low temperatures, the value of $r(t)$ has not considerable change in the timescale $t\ll\tau \approx {1}/{k_d}$, and  could be approximated as a fixed value $r_{0}$.

Note that in this estimation, we  only consider the deterministic term of the time evolution of \text{r}. This result has been confirmed through the simulation as well.

To study the dynamics of the system, we solve Eq. (\ref{eq:FP_omega_broken}) numerically by applying the Crank-Nicolson numerical method. Moreover, using the No-Flux boundary condition, we solve the Fokker-Planck equation for the system, taking into account different values of the relevant parameters in our scenario. Here, we consider the time evolution of the system for a time window of about 300 Myr. The initial probability density is assumed to be a Gaussian distribution in the enantiomeric excess $\omega$, with zero mean and variance 0.05.

\begin{figure*}[!htbp]
    \includegraphics[width=1 \textwidth]{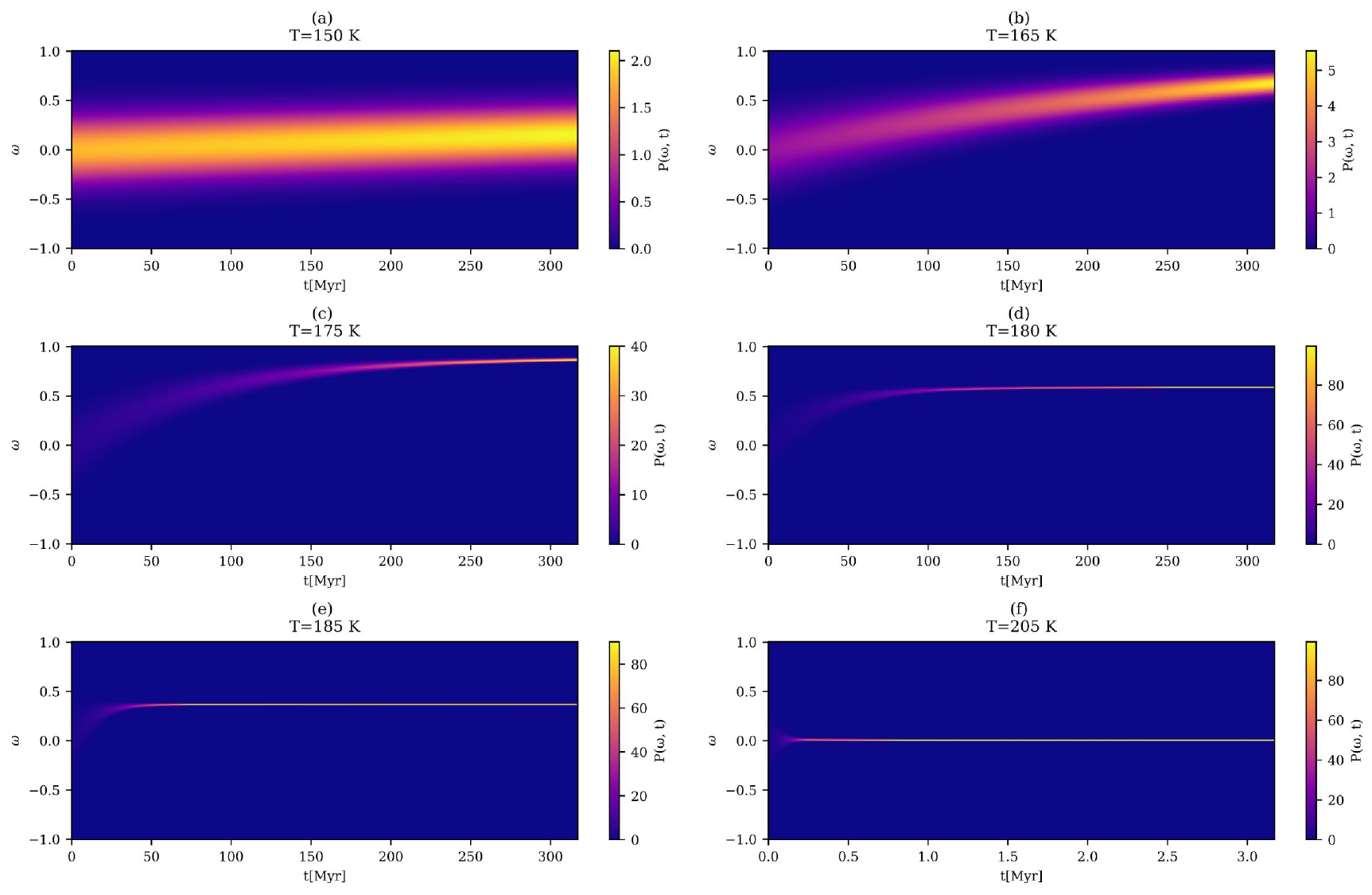}    % adjust width as needed
    \caption{Time evolution of the probability distribution of the system for different temperatures as labeled and for fixed parameters: V=$10^6$ $km^3$, L=1$0^{53}$ erg/s, r=0.01 AU, $G_{rac}$=26.30 kJ/mol. The vertical axis is enantiomeric excess, and the horizontal axis is time in Myr scale.}
    \label{Different Tl D}
\end{figure*}
\begin{figure}[!htbp]
\includegraphics[width=0.45\textwidth]{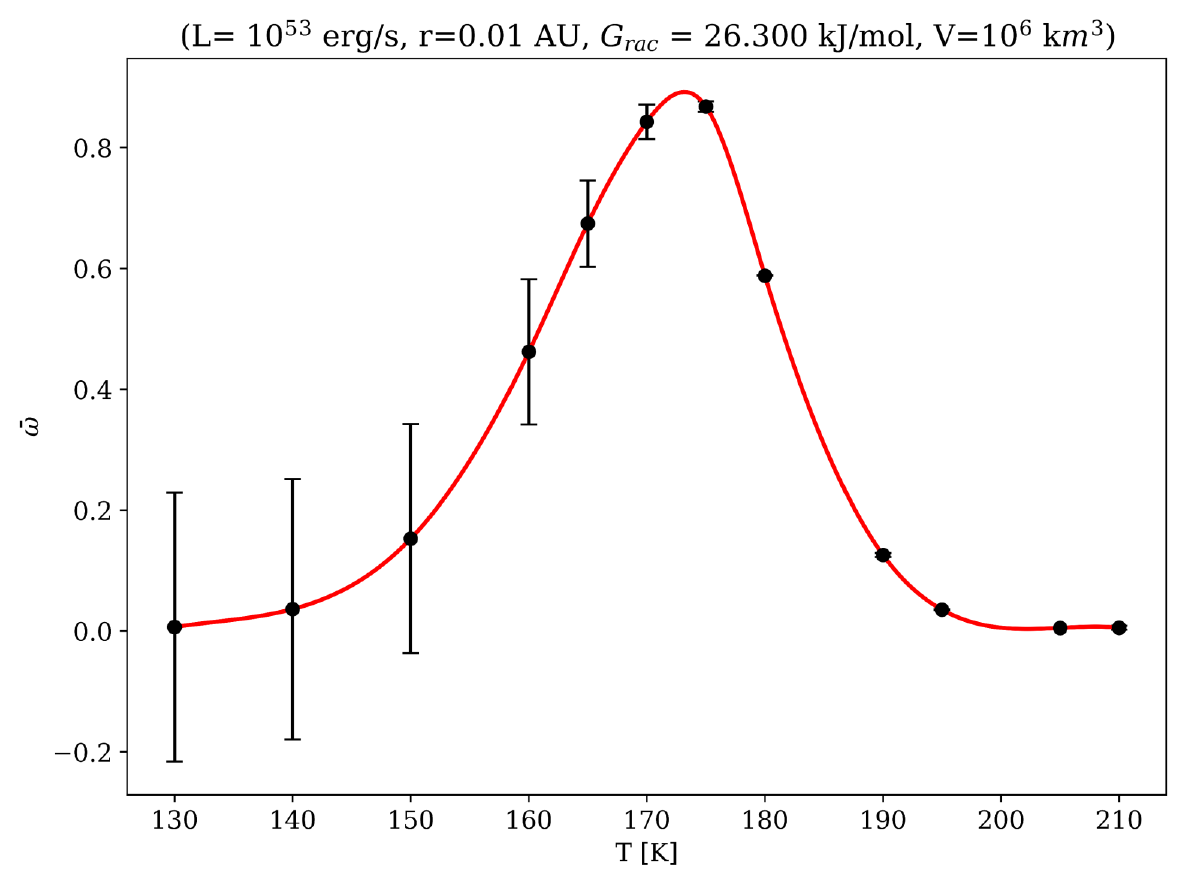}% Here is how to import EPS art
\caption{ Thermal evolution of the mean value of $\omega$ for the final probability distribution of the system at 300 Myr. The error bars show the standard deviation of the distribution at each temperature for fixed parameters : V=$10^6$ $km^3$, L=1$0^{53}$ erg/s, r=0.01 AU, $G_{rac}$=26.30 kJ/mol.}
\label{Omega}
\end{figure}

\begin{figure*}[!htbp]
\includegraphics[width=1\textwidth]{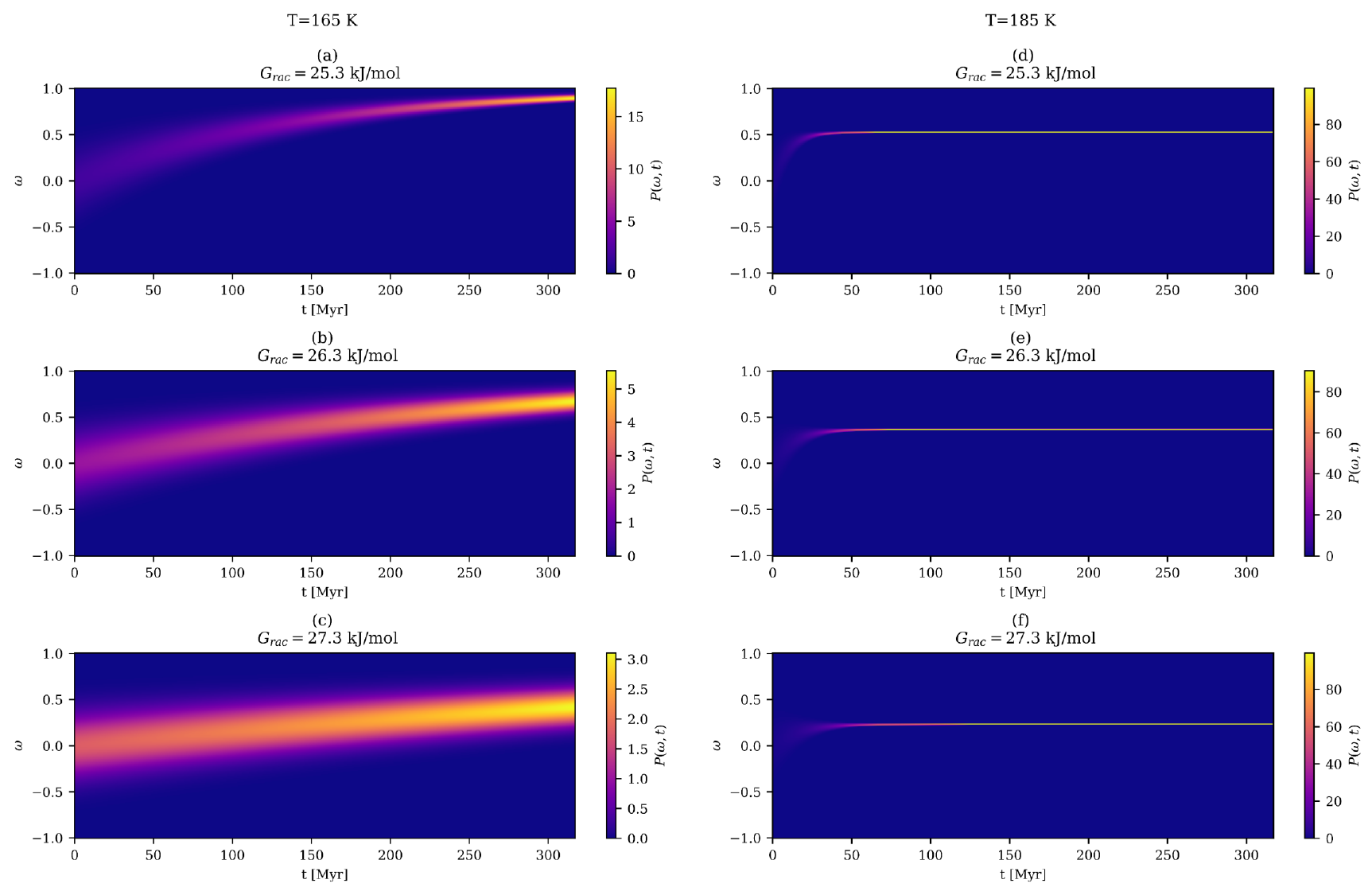}% Here is how to import EPS art
\caption{ Time evolution of the probability distribution of the system for different racemization activation energies as labeled. Other parameters are fixed at V=$10^6$ $km^3$, L=1$0^{53}$ erg/s, r=0.01 AU and T=165 K.}
\label{G Prob (165)}
\end{figure*}

Fig. \ref{Different Tl D} shows the evolution of the probability of the system for different temperatures as labeled, and the rest of parameters volume, (V=$10^6$ $km^3$), activation energy of racemization ($G_{rac} = 26.30 kJ/mol$), luminosity of the supernovea ($L=10^{53} erg/s$) and the distance of the IMC from the supernovea (r=0.01 AU) are assumed to be constant for the plots. Fig. \ref{Different Tl D} (a), (b), and (c) show  that in lower temperatures, the parity-violation effect, despite having the dominant contribution, cannot convey the system to a final steady state at 300 Myr, and by increasing the temperature  ($T>175 K$), $k_s$ enhances and the system reaches the steady state in this time scale. Fig. \ref{Different Tl D} (d), (e), and (f) show  that after a critical temperature ($T\approx 180 K$), the enantiomeric excess begins to reduce, reaching a racemic final state at around $T \approx 205 K$ (where the parity-violation effects are strongly suppressed). Since other chemical reactions' effects become comparable to $k_s$, there will be a competition between parity-violation and decay reaction. In addition, as the temperature increases, the system reaches the steady state sooner than at lower temperatures, and, as the deterministic term dominates the noise, the system ends in a sharp state, moving towards $\omega \approx 0$.

Fig. \ref{Different Tl D} implies that for higher temperature of the system, it is less likely to end up with a non-zero enantiomeric excess state; on the other hand as the temperature of the system decreases, non-zero enantiomeric excess states are more likely to happen, but lower than a certain temperature, it takes longer than 300 Myr as the estimated time duration for homochirality emergence (see Table. \ref{tab:racemization}) \cite{cline1996homochiral,cline2005physical}.

Fig. \ref{Omega} shows the thermal evolution of the mean value of $\omega$ for the final state of the system at 300 Myr. The error bars are computed from the standard deviation of the final state's distribution in Fig. \ref{Different Tl D}. The temperature affects the time scale of the system dynamics, for the lower temperatures, the system does not reach its stationary state in 300 Myr; however, for the higher temperatures, the system stabilizes in several Myr or less, and there is a small window of opportunity  to generate the homochiral state.

To investigate the influence of the racemization energy on the dynamical evolution of the system, we consider two selected temperatures, 165 K and 185 K, equivalent to the states in which the system does not reach its steady-state by 300 Myr and does, respectively. The temperature is fixed at T=165 K for the right column and 185 K for the left column of Fig. \ref{G Prob (165)} show the evolution of the probability of the system for different activation enrgies as labeled for time scale 300 Myr, and the rest of parameters, volume (V=$10^6$ $km^3$), luminosity of the supernovea ($L=10^{53} erg/s$) and the distance of the IMC from the supernovea (r=0.01 AU) are fixed. For T=165 K, when the racemization activation energy increases, the probability of the system lying in a higher enantiomeric excess state reduces, and the probability distributes in a wider range of $\omega$, while for T=185 K, the enantiomeric excess of the equilibrium state decreases and the system ends up in a delta-like distribution at a specific $\omega$.

Note that other parameters such as the supernova luminosity and the distance of IMC from the supernova can change the value of $k_s$, and consequently, the dynamics of the system. According to Eq. (\ref{e4}), $k_s$ is proportional to the supernova luminosity and the inverse square of the distance through $n_{\nu}$. Additionally, the volume of the system, which is contained in the noise amplitude, does not affect the system considerably for the typical size of IMCs.

\vspace{-0.5cm}

\section{\label{VI}{Conclusion}}

We introduce a model in which both effects of auto-catalysis and parity violation are combined to explain the origin of biomolecular homochirality. Our mechanism takes into account the stochastic effects from noise in the system, leading to a final homochiral state of the system. The parity violation effects regarding the interaction of supernova neutrinos with chiral molecules provide a preference for the system to choose one of the enantiomers. Considering the parity violation effect as part of the chemical scheme, the large flux of supernova neutrinos plays a crucial role in generating homochiral states. We show that for  a given set of reasonable parameters, the enantiomeric excess can be increased up to more than $10\%$, which is consistent with the evidence that we have from analysis of the meteorites. Our results show that the probability distribution is sensitive to the physical parameters of the system. Although increasing the temperature from a lower limit leads to chiral symmetry breaking and an almost homochiral state in a given time scale of about 300 Myr, by passing a critical temperature, other chemical reactions cause the symmetry restoration, resulting in a final racemic state at high temperatures.

In comparison with previous studies \cite{Blackmond2010CSH,Blackmond2004Nature, HiggsBlackmond2025} in which autocatalytic reactions are investigated as the only mechanism for the emergence of homochirality,  we propose a general scenario to produce homochirality from an astrophysical origin, and if it is a precursor for early life, this gives a chance to find extraterrestrial life. In contrast to the previous studies \cite{RN70}, we have tried to apply more realistic parameters, and also include reversible chemical reactions. 

Our model has the advantage of considering the parity violation interaction as a part of the chemical scheme, modifying both deterministic and stochastic terms rather than imposing a chiral selectivity only as a bias term in reaction constants (e.g., in \cite{Kondepudi1985,RN74,RN71} ), which leads to the crucial role of supernova neutrino flux to derive the system 
and triggering PVED between two enantiomeric states in IMC.
While the parity violation effects of supernova neutrinos are tiny, our model introduces an amplification mechanism based on the stochastic processes to obtain a significant impact at the final state. Although our scenario is based on some simple assumptions, it introduces a novel way to generate biomolecular chirality in IMC around a supernova from neutrino interactions, and it may shed light on the chiral puzzle of life.
\vspace{-0.4cm}
\section*{{Acknowledgments}}
The authors are grateful to Farshid Jafarpour, Hossein Farrokhpour, and Keivan Aghababaei Samani for insightful comments and discussions. The authors dedicate this work to the memory of Prof. Farhad Fazileh, an exceptional teacher and mentor whose legacy continues to inspire generations of students.
%\bibliography{apssamp}
\providecommand{\noopsort}[1]{}\providecommand{\singleletter}[1]{#1}%

\end{document}